\documentclass[aps,prb,twocolumn,groupedaddress,showpacs,showkeys]{revtex4}
\usepackage{graphicx}
\usepackage{subfigure}

\begin{document}

\title{Current localization and Joule self-heating effects in Cr doped
Nd$_{0.5}$Ca$_{0.5}$MnO$_3$ manganites}

\author{A. S. Carneiro and R. F. Jardim}\email{rjardim@if.usp.br}
\affiliation{Instituto de F\'{i}sica, Universidade de S\~ao Paulo,
CP 66318, 05315-970, S\~ao Paulo, SP, Brazil}
\author{ F. C. Fonseca}
\affiliation{Instituto de Pesquisas Energ\'eticas e Nucleares, CP
11049, 05422-970, S\~ao Paulo, SP, Brazil}

\date{\today}

\begin{abstract}
The effects of dc excitation current on the current-voltage curves
of polycrystalline samples of
Nd$_{0.5}$Ca$_{0.5}$Mn$_{0.96}$Cr$_{0.04}$O$_{3}$ were
investigated. The experimental results show that an abrupt jump of
the voltage is concomitant with a huge increase in the temperature
of the sample. A simple model and estimates for Joule self-heating
effects support the experimental data. Moreover, the data strongly
suggest that both the current localization in the metallic paths
and local Joule self-heating effects are essential ingredients to
understand the current-induced phase transition in phase-separated
manganites.
\end{abstract}

\maketitle

It has been reported that phase control in manganites can be
achieved by both the appropriate chemical substitutions,\cite{1}
and by application of external stimuli such as pressure
$P$,\cite{2} magnetic field $H$,\cite{1} high-power laser
irradiation,\cite{3} and electric fields $E$.\cite{4} However,
while both $H$ and $P$ change the bulk properties of the colossal
magnetoresistance (CMR) manganites, the influence of light,
X-rays, and electric current $I$ may result in a rather localized
phase transition.\cite{5} In particular, effects induced by
applied $E$ have been studied in both the charge/orbital-ordered
(CO-OO) and the ferromagnetic-metallic (FMM) states in single
crystals and polycrystalline samples of CMR
manganites.\cite{4,5,6,7,8,9,10} In these compounds, a drop in the
electrical resistivity $\rho(T)$ above a threshold value of the
applied dc $E$ (or dc current $I$) is generally
observed.\cite{6,7} This $E$-driven (or $I$-driven) insulator to
metal transition MI is close related to the occurrence of a
nonlinear conductivity.\cite{4,6} The nonlinearity of the $I$-$V$
curves, observed for $T$$<$$T_{CO}$, is thus explained by
considering a percolation process due to the melting of the CO
insulating (COI) phase into the metallic one. This interpretation
is based on the opening of metallic-filament paths in a COI
matrix.\cite{4,5,6}

Besides the common observation of nonlinearity in $I$-$V$ curves,
this scenario is still not clear and other physical mechanisms
have been proposed as the depinning of the CO-OO state upon $E$
application,\cite{7} the change in the orientation of the OO in
the insulating state upon application of large $E$,\cite{10} and
an $E$-induced switching of the directional order of the OO states
of \textit{e}$_g$ electrons.\cite{11} Results of $I$-$V$ curves
under $H$ also indicated an increase of the magnetization at the
same threshold $V$ where $I$ rises abruptly.\cite{9} It was argued
that Joule heating effects were insignificant and the formation of
a less-resistive mixed state, but with higher magnetization than
the initial CO state, was suggested.\cite{9} More recently, the
nonlinear conduction was described by considering a model for
charge-density-wave (CDW) motion.\cite{12}

The proposed scenarios for the $I$ (or $E$)-induced phase
transition usually consider mechanisms related to the interplay
between charge and spin degrees of freedom.\cite{7,8,9,10,11,12}
However, based on magneto-optical imaging of local magnetization
and $I$ distribution, a different view has been recently
proposed.\cite{13} It was suggested that application of high
density $I$ reduces the metallic channels due to local Joule
self-heating effects. Increasing $I$ changes the metallic paths
into insulating regions, further localizing the inhomogeneous
current flow, and propagating the Joule self-heating. This process
results in a rapid collapse of the metallic channels, which is
related to an abrupt jump in the $I$-$V$ curves at a well-defined
$I$ threshold $I_{T}$.\cite{13}
\begin{figure}[htp]
\centering
\includegraphics[width=7cm]{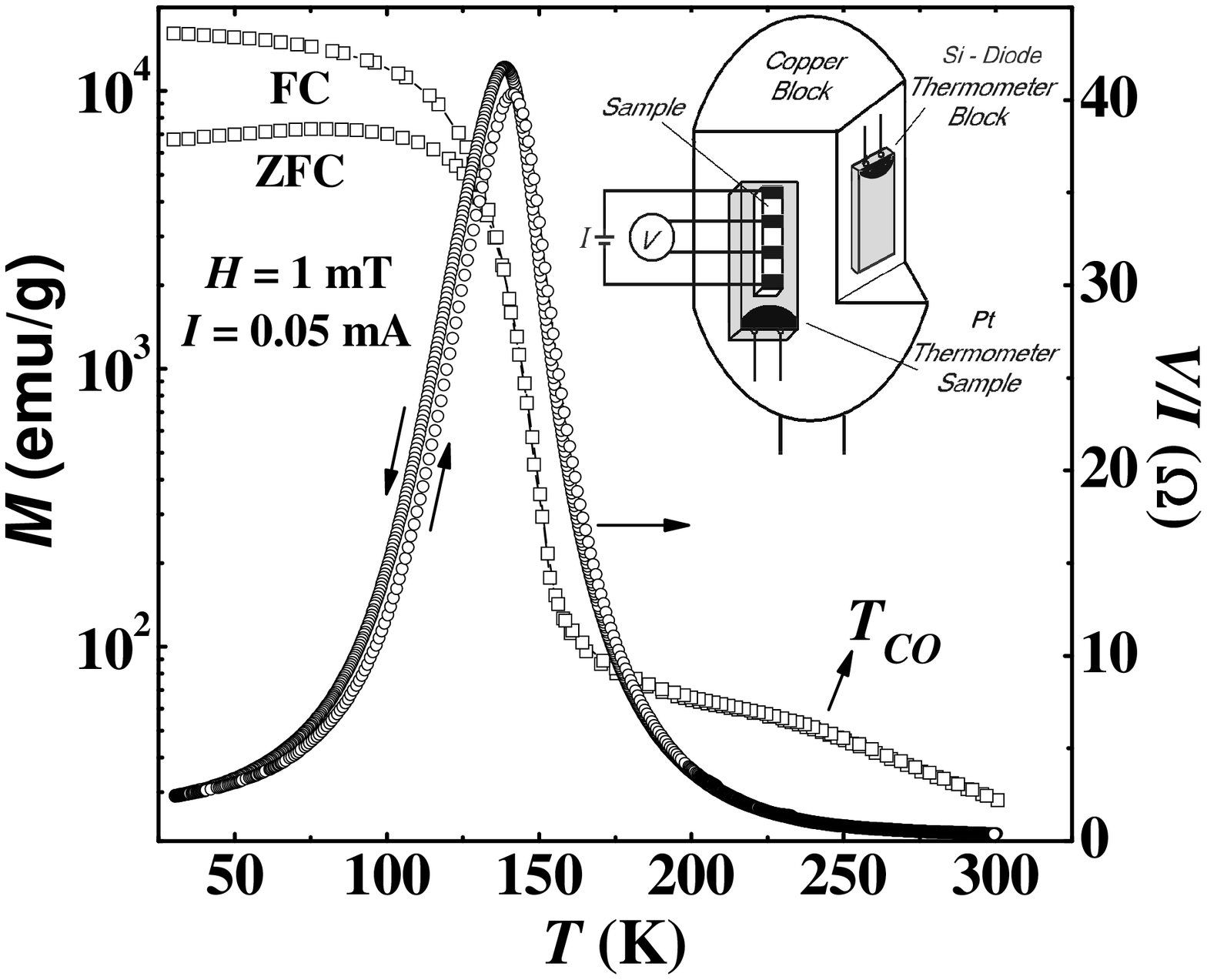}
\caption{\label{fig1}$V/I(T)$ (right axis) and $M(T)$ (left axis)
of Nd$_{0.5}$Ca$_{0.5}$Mn$_{0.96}$Cr$_{0.04}$O$_{3}$. The inset
displays the copper block sample holder experimental setup.
Warming and cooling cycles in $V/I(T)$ measurements ($I$=0.05 mA)
are indicated by arrows.}
\end{figure}

We report the occurrence of abrupt jumps/drops on the $V$ across
the phase-separated manganite
Nd$_{0.5}$Ca$_{0.5}$Mn$_{0.96}$Cr$_{0.04}$O$_{3}$ (Cr-NCMO). By
using an appropriated experimental setup we have observed that the
occurrence of nonlinearity in $I$-$V$ curves is a precursor of a
drastic rise in the temperature of the sample $T_S$. A model to
account for the increase in $T_S$ on increasing $I$ supports the
experimental data. Moreover, based on the temperature dependence
of the specific heat $C_{p}$, an estimate of the energy
dissipation of the material further confirms the importance of
Joule heating effects. The results allowed us to propose that
Joule self-heating effects associated with localized conduction
are essential ingredients to understand the abrupt $I$-induced
phase transition in phase-separated manganites.

Polycrystalline samples of Cr-NCMO were prepared by solid state
reaction.\cite{14} Four-wire dc $\rho(T)$ and $I$-$V$ measurements
were performed between 30 and 300 K. Four gold contact pads were
deposited on parallelepiped-shaped (typically of length $l$=0.60
cm, width $w$=0.145 cm, and thickness $t$=0.075 cm) samples to
obtain excellent electric contacts by using Ag epoxy. In $I$-$V$
measurements, the current sweeps between -100$\leq$$I$$\leq$100 mA
were carried out in 0.5 mA steps. To avoid any memory effects,
after each measurement the sample was heated to 300 K, cooled down
to 30 K, and then heated to the desired measuring $T$. The
experimental setup for the electrical measurements uses a cold
finger connected to a closed-cycle helium refrigerator. The
temperature of the system $T$ is monitored with a silicon diode,
positioned near the sample, and mounted on the copper block. The
temperature of the sample $T_S$ is measured by a Pt thermometer
placed on the copper block. The sample stands over the Pt
thermometer and is fixed to it by using a thin thermal conducting
silicon grease layer. The inset in Fig.\ref{fig1} displays a
schematic draw of the sample holder setup. Magnetization $M(T)$
measurements were performed in a SQUID magnetometer under applied
magnetic field of 1 mT in both FC and ZFC modes. $C_{p}$
measurements in Cr-NCMO crystals were performed in a Quantum
Design PPMS apparatus.
\begin{figure}[htp]
\centering
\includegraphics[width=7.5cm]{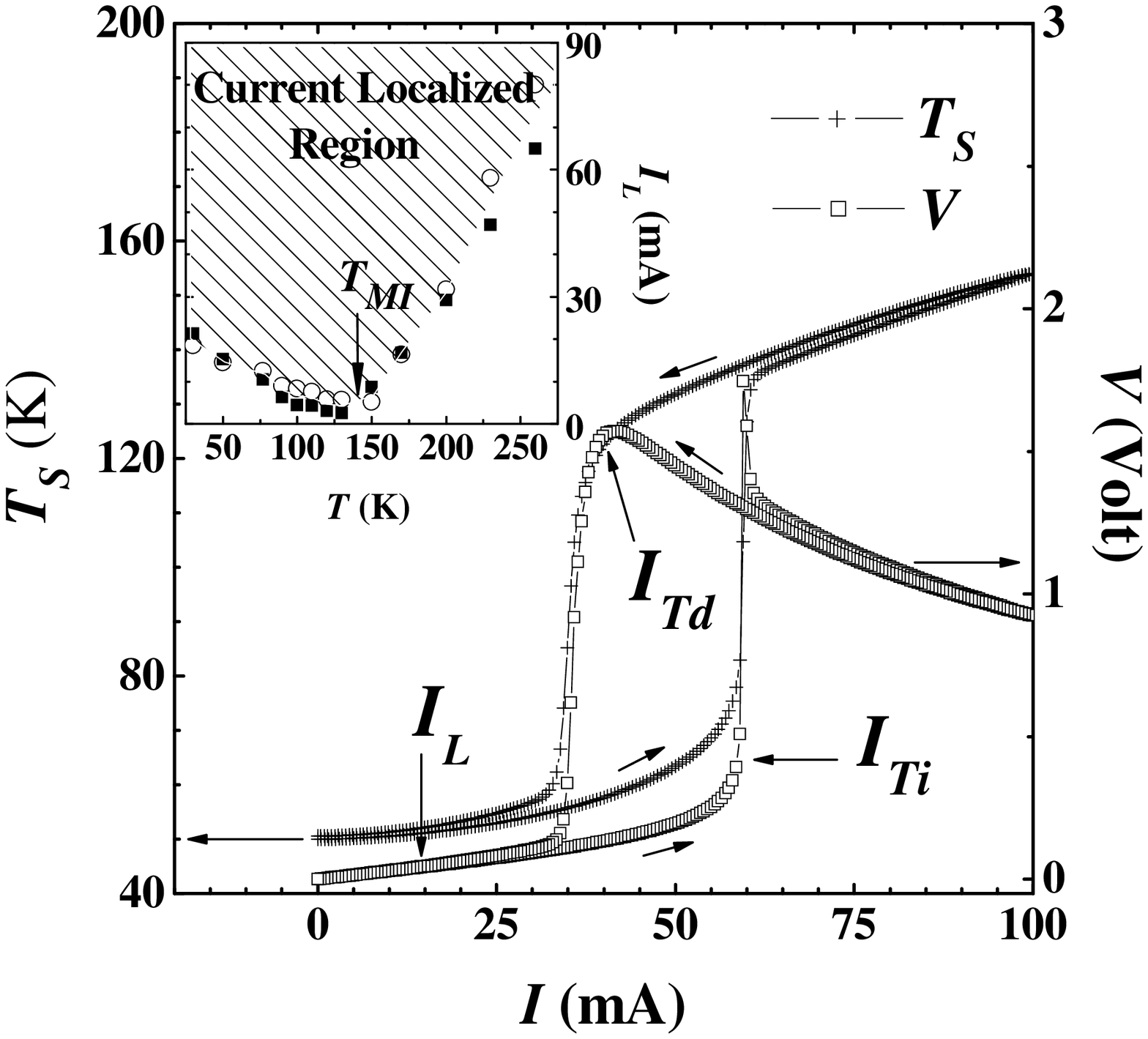}
\caption{\label{fig2} $I$-$V$ curve at 50 K (right, open squares)
and the corresponding temperature rise of the sample $T_S$ (left,
crosses) vs. $I$. The inset shows the temperature dependence of
$I_{L}$. Arrows denote increasing and decreasing $I$.}
\end{figure}

The compound Nd$_{0.5}$Ca$_{0.5}$MnO$_{3}$ has a CO
antiferromagnetic-insulator (AFI) ground state below
$T_{CO}$$\sim$240 K where a partial OO and magnetic correlations
of short range are observed.\cite{1,15} At low $T$, the OO
increases and a long range CE-type AFI state is established below
$T_{N}$$\sim$170 K. Substitutions on the Mn site are an effective
way to gradually modify the CO state.\cite{1} The material under
investigation Nd$_{0.5}$Ca$_{0.5}$Mn$_{0.96}$Cr$_{0.04}$O$_{3}$ is
well studied and develops a FMM phase at low temperatures, as
shown in Fig.\ref{fig1}.\cite{1,15} The transition to the CO
regime is inferred from a subtle change of $d\rho/dT$ and a small
cusp in the $M(T)$ data at $T_{CO}$$\sim$240 K.\cite{1,15} With
decreasing $T$, a MI transition, closely related to the FM
transition, occurs at $T_{MI}$$\sim$140 K when $I$=0.05 mA. The
thermal irreversibility in both $M(T)$ and $\rho(T)$ data indicate
a coexistence of CO/OO and FMM phases at low
temperatures.\cite{14,16} Indeed, under Cr-doping and for
$T$$<$$T_{MI}$, these compounds are comprised of a fine mixture of
20-30 nm domains of the FMM phase embedded in the CO/OO
matrix.\cite{16}

Fig.\ref{fig2} shows a typical $I$-$V$ curve ($T$=50 K) and the
temperature rise of the sample $T_S$ as $I$ increases. The data
reveal a clear correspondence of both $V$ and $T_S$ curves against
$I$. At low $I$, the rising in $T_S$ is smooth and the $I$-$V$
data is here considered linear up to an $I$ value
$I_{L}$$\sim$14.5 mA. $I_{L}$ is defined as the current in which
the temperature measured by the Pt-thermometer increases 1 K. As
$I$ evolves, the $I$-$V$ data loose the linearity and $T_S$ rises
progressively. Further rising in $I$ leads to a switch to a much
higher $V$/$I$ value, as inferred from a steep increase in $V$ at
$I_{Ti}$$\sim$59 mA ($\sim$5.4 A/cm$^{2}$). This $I$-induced phase
transition is accompanied by a remarkable increase in $T_S$ from
$\sim$70 to $\sim$135 K in a narrow range of $I$ while $V$
displays a spike-like maximum. With further increasing in $I$, $V$
decreases monotonically, i. e., the $I$-$V$ curves exhibit a
negative differential resistance ($dV/dI$$<$0) up to $I$=100
mA.\cite{6} This feature is also reproduced in $T_S$ that rises
continuously, reaching $\sim$155 K for $I$=100 mA, while the
temperature of the copper block varied $\sim$1.6 K.

The $I$-induced phase transition at $I_{Ti}$ is useful to separate
two well defined regions in dynamical $I$-$V$ curves: below and
above $I_{Ti}$, where the material can be considered as a metal or
insulator, respectively. The decreasing branch of $I$-$V$ curves
shows a pronounced irreversibility in $I$, preserving the more
insulating state down to a lower value of the threshold current
$I_{Td}$($T$=50 K)$\sim$40 mA. At this excitation current, $V$
drops rapidly, and the initial $I$-$V$ curve is recovered with
further decreasing $I$. Again, $T_S$ follows essentially the same
trend: it decreases from $T_S$$\sim$150 K but it is kept above 100
K while the more insulating state is preserved. Accordingly, close
to $I_{Td}$($T$=50 K)$\sim$40 mA, $T_S$ drops rapidly to $\sim$55
K and continuously decreases towards to the initial value (50 K).
\begin{figure}[htp]
\centering
\includegraphics[width=7cm]{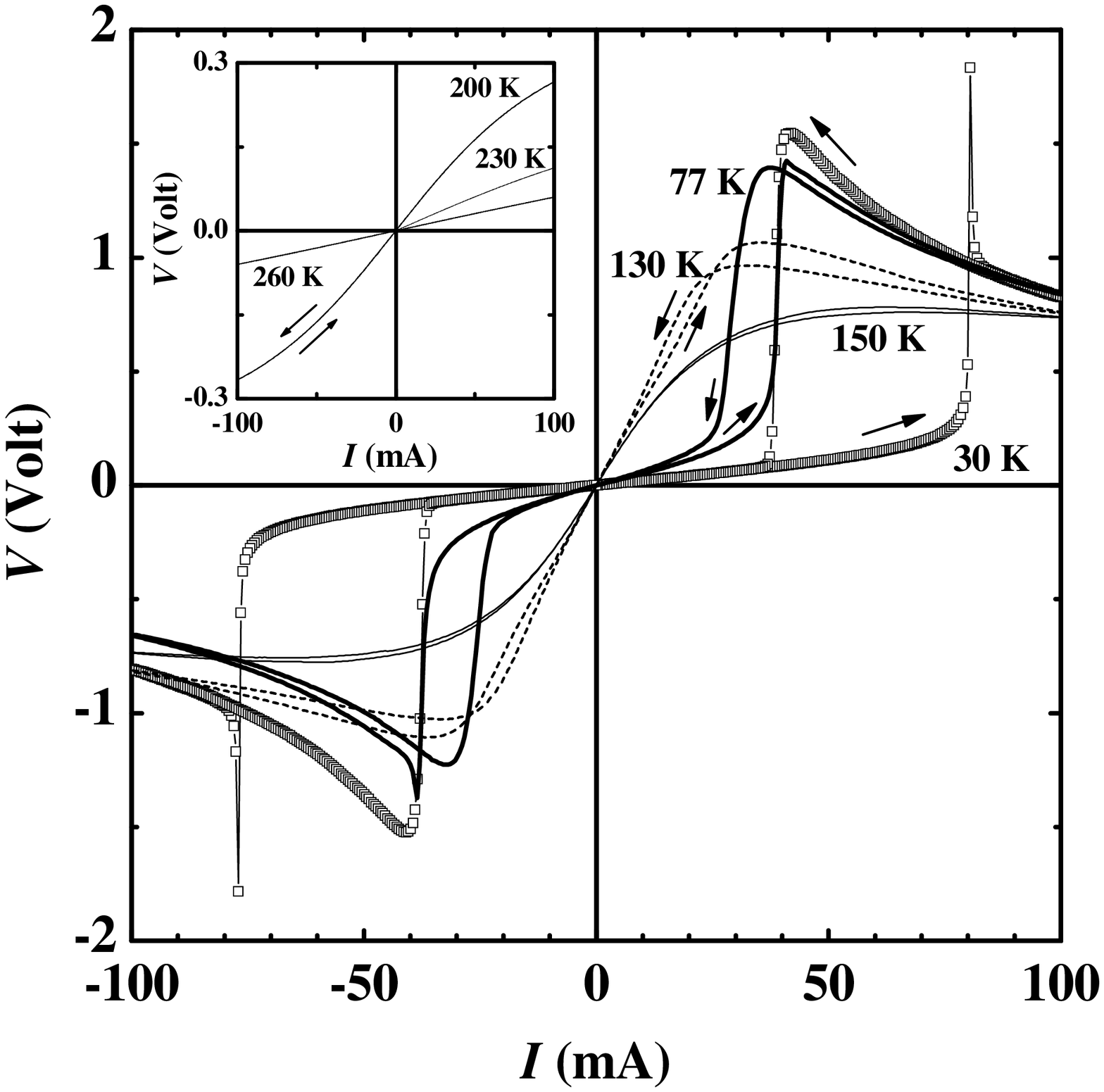}
\caption{\label{fig3} $I$-$V$ curves of the Cr-NCMO compound at
several temperatures. The inset displays the data in the vicinity
of $T_{CO}$. The arrows point to the increasing and decreasing
values of $I$.}
\end{figure}

Due to a thin conducting grease layer separating the
Pt-thermometer and the sample, a temperature gradient $T_G$ in
this region is expected to occur. An estimate of $T_G$ is made by
comparing values of $V$/$I$ in $I$-$V$ curves and considering the
$V$/$I$ vs. $T$ data of Fig.\ref{fig1}. From Fig.\ref{fig2}, the
spike-like maximum in $V$/$I$$\sim$29 $\Omega$ ($I$$\sim$60 mA)
corresponds to an average temperature reached by the sample of
$\sim$152 K (see Fig.\ref{fig1}), a value compatible with the
transition from metallic to insulating behavior but slightly
higher than $T_S$$\sim$135 K measured by the Pt-thermometer. Such
a difference $T_G$$\sim$17 K is certainly related to the thermal
coupling between sample and the Pt-thermometer but not only. A
temperature gradient between the sample surface and the bottom is
also expected since the thermal conductance in manganites is
rather low.\cite{17} In addition, the estimated average sample
temperature extracted from the $V$/$I$ data of Fig.\ref{fig1},
taken at constant and rather low value of $I$=0.05 mA, must be
also considered. For such low values of $I$ a homogeneous
distribution of $I$ within the sample is expect, a feature hardly
believed to occur at $I$$\sim$60 mA.

A set of $I$-$V$ curves taken at different $T$ is displayed in
Fig.\ref{fig3}. The most prominent feature here is the sharp jump
in $V$ at $I$$\sim$$I_{Ti}$, which decreases appreciably with
increasing $T$ and is barely identified for $T$$>$$T_{MI}$ (inset
of Fig.\ref{fig3}). Increasing $T$ also results in a less
pronounced irreversibility of the $I$-$V$ curves for
$T$$<$$T_{MI}$ and, for $T$$>$$T_{MI}$, they exhibit a reversible
behavior.\cite{18} This indicates that the irreversibility in
$I$-$V$ curves is close related to the phase competition in the
phase-separated manganite, being larger when the FMM phase is
robust, and further evidences the localized electric conduction in
the FMM phase. The sweep to negative $I$ values leads to an
antisymmetric $I$-$V$ behavior of the curves.

From $I$-$V$ curves at different $T$, values of $I_{L}$ were
computed and are shown in the inset of Fig.\ref{fig2}. Such a
diagram is useful for separating regions where current
localization takes place in these materials. The decrease of
$I_{L}$ with increasing $T$$\leq$$T_{MI}$ reflects the reduction
of the relative volume fraction (VF) of the FMM phase, which
coexist with the CO/AFI.\cite{1,14,16} Such a reduction is
consistent with a more localized $I$ distribution across the
material and provides an explanation for the so-called $I$-induced
change in $T_{MI}$.\cite{19,20} A considerable increase of $I_{L}$
for $T$$>$$T_{MI}$ suggests a much more homogenous distribution of
$I$ across the sample in the insulating phase.\cite{13,20}
However, the current localization at high applied $I$ and
$T$$>$$T_{MI}$ is probably related to the inhomogeneous nature of
the CO-state, in which different structural phases were found to
coexist.\cite{15,20a} These observations suggest that both the
temperature dependence of $I_{L}$ and the abrupt jump in $V$ are
intimately related to the VF of the metallic phase within the
material which is believed to vanish when the average temperature
of the sample is $T$$\gtrsim$$T_{MI}$.
\begin{figure}[htp]
\centering
\includegraphics[width=7cm]{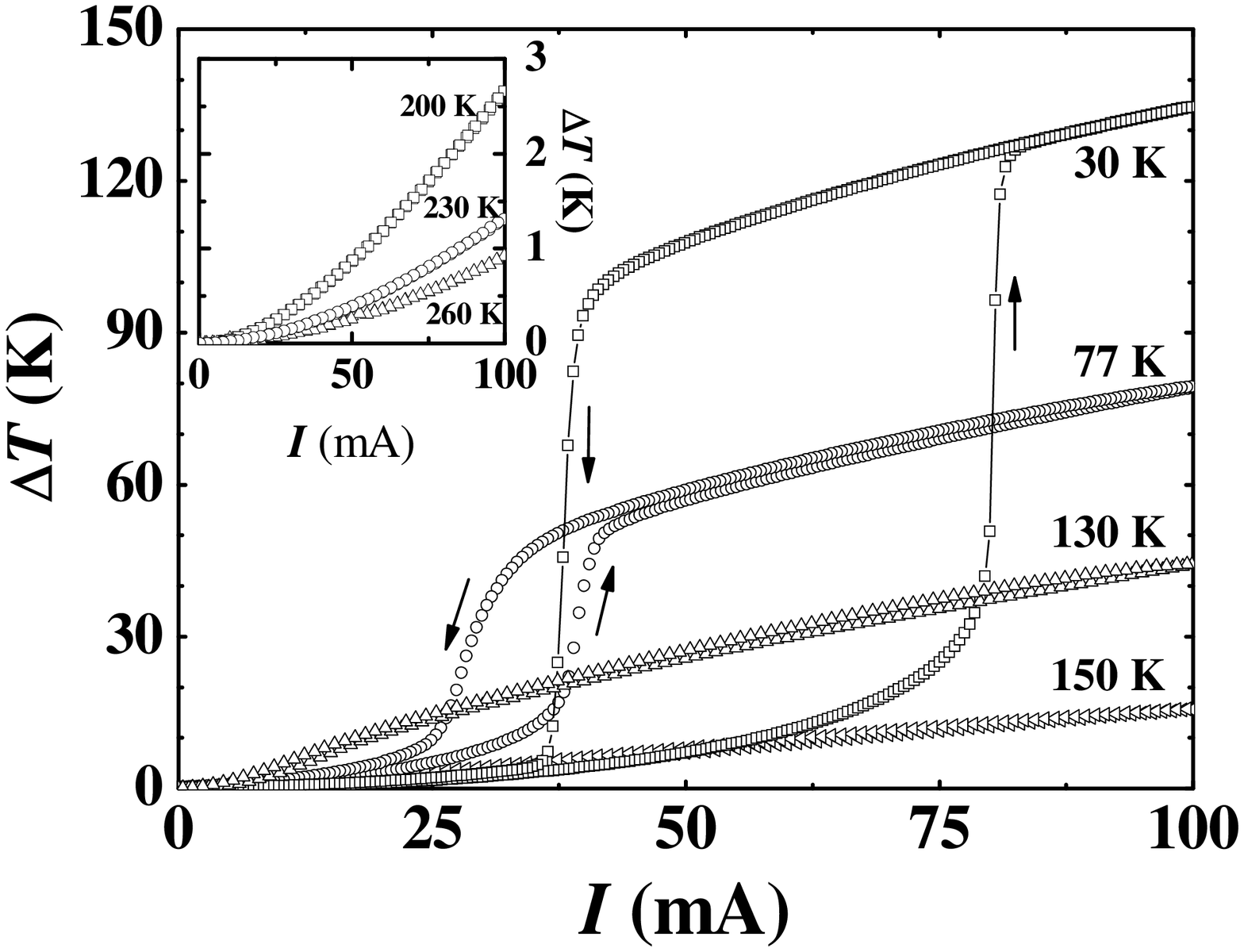}
\caption{\label{fig4} Temperature rise of the sample $\Delta
T$=$T_S$-$T$ vs. $I$. The copper-block sample holder measuring
temperatures $T$ for each $I$-$V$ curve are indicated. The inset
shows $\Delta T$ close to $T_{CO}$. The arrows point to the
increasing and decreasing values of $I$.}
\end{figure}

We have carefully checked changes in the temperature of the copper
block in all $I$-$V$ curves. The data show a maximum temperature
rise of the block of $\sim$1.9 K at $T$=30 K, when the sample is
subjected to $I$=100 mA and $T_S$$\sim$165 K. Thus, the
temperature rise of the sample with respect to the copper block
has been defined as $\Delta T$= $T_S$-$T$ and typical data of
$\Delta T$ vs. $I$ are shown in Fig.\ref{fig4}. At $T$=30 K, a
huge rise of $\sim$100 K in $\Delta T$, at essentially the same
$I$$\sim$80 mA where $V$ increases drastically, is observed. The
average temperature reached by the sample $T_S$ just after the
jump was measured to be $\sim$155 K, a value higher than
$T_{MI}$$\sim$140 K. Such a rise in $T_S$ is certainly caused by
Joule self-heating effects, a process that continues up to $I$=100
mA, where $T_S$$\sim$165 K. When the sample was subjected to a
power dissipation of $\sim$0.1 W ($I$=100 mA), the highest value
of $\Delta T$$\sim$135 K was observed for $T$=30 K. Again, a
temperature gradient $T_G$$\sim$15 K, at $I$=100 mA, was
estimated, further indicating a current localization, as already
mentioned. Such a $T_G$ of the experimental arrangement is more
pronounced at high $I$ and $T$$\leq$$T_{MI}$, indicating that the
local temperature of the sample is significantly altered by the
current localization.

The behavior of $\Delta T$ also displays irreversibility in curves
taken at low temperatures, a feature that vanishes for
$T$$\geq$$T_{MI}$. Increasing $T$ results in smaller values of
$\Delta T$, being $\Delta T$$\sim$2 K for $I$=100 mA at
$T$$\sim$$T_{CO}$ (inset of Fig.\ref{fig4}). Therefore, the
experimental data suggest that changes in $I$-$V$ curves are close
related to Joule self-heating effects, to the VF of the FMM phase,
and to the localized distribution of $I$ in the metallic phase.
\begin{figure}[htp]
\centering
\includegraphics[width=7cm]{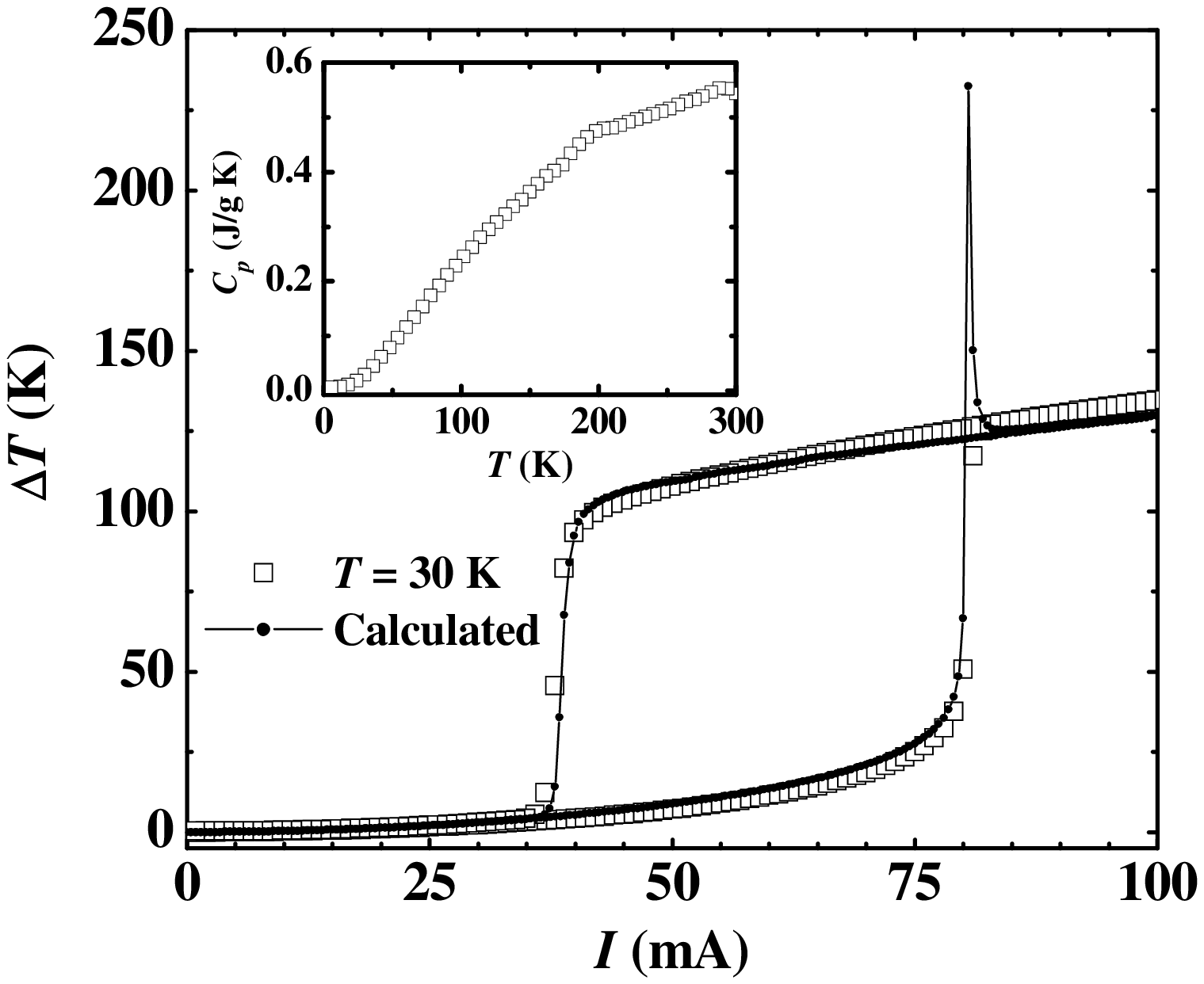}
\caption{\label{fig5} The $I$ vs. $\Delta T$ curve at 30 K (open squares) and the
calculated Joule-self heating effect (solid circles and line). The inset displays the
$C_{p}(T)$ data of a single crystal specimen.}
\end{figure}

To support the experimental findings, the expected increase in
$T_S$ due solely to Joule self-heating effects was calculated by
considering the heat dissipation in the sample under applied $I$
and the heat conduction by the grease placed between the sample
and the Pt thermometer.\cite{21} Since $T$ was found to be nearly
constant during the experiment, the temperature rise of the sample
$\Delta T(T)$ can be approximated by $P/\kappa_{g}$($A/d$), where
$P$ is the power dissipated in the sample, and $\kappa_{g}$ is the
thermal conductivity of the grease with area $A$ and thickness
$d$. By using the experimental $R$=$V$/$I$ data and
$P$=$I^2$$\rho$*($l/wt$), $\Delta T(T,I)$=$I^{2} \rho$($T$+$\Delta
T$)/$w$$t$$\kappa_{g}$*($ld/A$), where $w$, $l$, and $t$ were
already defined. Considering that $\kappa_{g}$$\sim$0.16
Wm$^{-1}$K$^{-1}$ is nearly temperature independent in the $T$
range investigated (30-160 K),\cite{22} an example of the
calculated $\Delta T$($T$=30 K,$I$) for a typical sample is shown
in Fig.\ref{fig5}. The good agreement between the calculated and
the measured $\Delta T(T)$ confirms that Joule self-heating
provides an excellent description for the temperature rise of the
sample. Small deviations in the high $I$ region are mainly related
to: (i) the transition to the insulating regime where the VF of
the FMM is low; (ii) changes in $T$ that is assumed to be
constant; (iii) a thermal gradient within the sample; and (iv) a
more homogeneous distribution of $I$ across the sample (see inset
of Fig.\ref{fig2}).\cite{5,13} Furthermore, the $C_{p}(T)$ data of
a Cr-NCMO single crystal specimen (inset Fig.\ref{fig5}) allowed
us to estimate the energy required to heat the sample at a given
temperature and to compare with the one due to Joule self-heating
effects. By using the $C_{p}$ data, the energy required to heat a
Cr-NCMO sample from 30 to 140 K was estimated to be $\sim$85 mJ.
Accordingly, the energy dissipated by $I_{Ti}$($T$=30 K) was found
to be $\sim$80 mJ, corresponding to a $T_S$$\sim$140 K, in
excellent agreement with the data shown in Fig.\ref{fig5}). This
result supports the experimental data, and lends credence to the
relationship between Joule self-heating effects and the
temperature rise of the sample.

In summary, the combined experimental results indicate that the
abrupt increase of $T_S$ can be doubtless ascribed to the heat
generated by the $I$ flow through localized pathways in the FMM
phase. Under relatively low $I$, the weaker metallic regions
become insulating, increasing the current density in the remaining
FMM paths.\cite{5} This is reflected in both the initial rise of
$T_S$ and changes in the $I$-$V$ behavior. As $I$ evolves, the
weaker metallic paths are progressively reduced and the Joule
self-heating effect increases, culminating in a rapid collapse of
the FMM phase.\cite{5} This is mirrored in the observed jumps in
$V$ and the dramatic rise of $T_S$. The Joule self-heating is
strong enough to promote the sample to the insulating phase
($T_S$$>$$T_{MI}$), resulting in a less appreciable self-heating
due to a much more homogeneous distribution of $I$.\cite{13,20}
The negative differential resistance at high $I$ values (Figs.2
and 3) is a consequence of the decrease in $V$/$I$($T$$>$$T_{MI}$)
as $T_S$ still rises with increasing $I$.

This work was supported by the Brazilian agency FAPESP under Grant
No. 99/10798-0, 01/01454-8, and 01/04231-0. R. F. J. and F. C. F.
are CNPq (Brazil) fellows under Grant No. 303272/2004-0 and
301661/2004-9, respectively.


\begin{references}
\bibitem{18} The irreversible behavior of both $V$ and $T_S$ vs. $I$ curves below $T_{MI}$ are further evidence of the phase-separation in this manganite. In this
$T$ range both $I_{Ti}$ and $I_{Td}$, along with the difference $I_{Ti}$-$I_{Td}$,
decrease with increasing $T$ and approach zero at $T$$\sim$$T_{MI}$.
\end{references}
\end{document}